\documentclass[11pt,aaspp4]{aastex}

\slugcomment{\it 2012ApJS..200....9G}

\usepackage{graphicx} 
\usepackage{amssymb}
\usepackage{amsmath}
\usepackage{anysize}
\usepackage{rotating}
\usepackage{multirow}
\usepackage{float}

\begin{document}

\title{The Advanced Camera for Surveys General Catalog: \\
Structural Parameters for Approximately Half a Million Galaxies}

\author{Roger L. Griffith\altaffilmark{1}, Michael C. Cooper\altaffilmark{2},\altaffilmark{\dag}, Jeffrey A. Newman\altaffilmark{3}, Leonidas A. Moustakas\altaffilmark{4}, Daniel Stern\altaffilmark{4}, Julia M. Comerford\altaffilmark{5}, Marc Davis\altaffilmark{6}, Jennifer M. Lotz\altaffilmark{7}, Marco Barden\altaffilmark{8}, Christopher J. Conselice\altaffilmark{9}, Peter L. Capak\altaffilmark{10}, S. M. Faber\altaffilmark{11}, J. Davy Kirkpatrick\altaffilmark{1}, Anton M. Koekemoer\altaffilmark{7}, David C. Koo\altaffilmark{11}, Kai G. Noeske\altaffilmark{12}, Nick Scoville\altaffilmark{10}, Kartik Sheth\altaffilmark{10}, Patrick Shopbell\altaffilmark{10}, Christopher N. A. Willmer\altaffilmark{13}, Benjamin Weiner\altaffilmark{13}}

\altaffiltext{1}{Infrared Processing and Analysis Center, California Institute of Technology, Pasadena, CA 91125}

\altaffiltext{2}{Center for Galaxy Evolution, Department of Physics and Astronomy, University of California, Irvine, 4129 Frederick Reines
Hall, Irvine, CA 92697}

 \altaffiltext{\dag}{Hubble Fellow}

\altaffiltext{3}{Pittsburgh Particle Physics, Astrophysics, and Cosmology Center, Department of Physics and Astronomy, University of Pittsburgh,
Pittsburgh, PA 15260}

\altaffiltext{4}{ Jet Propulsion Laboratory, California Institute of Technology, 4800 Oak Grove Dr, Pasadena, CA 91109}

\altaffiltext{5}{Astronomy Department, University of Texas at Austin, Austin, TX, 78712}

\altaffiltext{6}{Department of Astronomy, University of California, Berkeley, Hearst Field Annex B, Berkeley, CA 94720}

\altaffiltext{7}{Space Telescope Science Institute, 3700 San Martin Dr., Baltimore, MD, 21218}

\altaffiltext{8}{Institute of Astro- and Particle Physics, University of Innsbruck, Technikerstr. 25, 6020 Innsbruck, Austria}

\altaffiltext{9}{University of Nottingham, School of Physics \& Astronomy, Nottingham}

\altaffiltext{10}{Spitzer Science Centre, 314-6 California Institute of Technology, 1200 E. California Blvd., Pasadena, CA, 91125}

\altaffiltext{11}{UCO/Lick Observatory, University of California}

\altaffiltext{12}{Keck Foundation Fellow; Harvard-Smithsonian Center for Astrophysics, Cambridge, MA,U.S.A.}

\altaffiltext{13}{Steward Observatory, University of Arizona, 933 North Cherry Avenue, Tucson, AZ 85721}


\begin{abstract}{ We present the Advanced Camera for Surveys General Catalog (ACS-GC), a photometric and morphological database using publicly available data obtained with the Advanced Camera for Surveys (ACS) instrument on the {\it{Hubble Space Telescope}}. The goal of the ACS-GC database is to provide a large statistical sample of galaxies with reliable structural and distance measurements to probe the evolution of galaxies over a wide range of look-back times. The ACS-GC includes approximately 470,000 astronomical sources (stars + galaxies) derived from the AEGIS, COSMOS, GEMS, and GOODS surveys.  G\textsc{alapagos} was used to construct photometric (SE\textsc{xtractor}) and morphological (G\textsc{alfit}) catalogs.  The analysis assumes a single S\'ersic model for each object to derive quantitative structural parameters.  We include publicly available redshifts from the DEEP2, COMBO-17, TKRS, PEARS, ACES, CFHTLS, and zCOSMOS surveys to supply redshifts (spectroscopic and photometric) for a considerable fraction ($\sim$74\%) of the imaging sample.  The ACS-GC includes color postage stamps, G\textsc{alfit} residual images, and photometry, structural parameters, and redshifts combined into a single catalog.}

\end{abstract}

\section{Introduction}

The detailed study of galaxy evolution began with the pioneering work of Edwin Hubble at Mt.\ Wilson Observatory in the 1920's. Hubble pioneered the investigation of galaxy properties by classifying galaxies according to their morphological structure, leading to the Hubble sequence of galaxies \citep{hub26}. In the local universe, the Hubble sequence is well defined and widely used; however, as one goes back in distance and cosmic time, morphological classification becomes an increasingly difficult problem. The advent and rapid growth of CCD technology within the past 30 years has allowed astronomers to image and catalog galaxies that were inaccessible in previous studies. In order to build a deep, comprehensive and coherent theory on galaxy evolution, {\it{complete}} samples of galaxies spanning a wide range of redshifts and look-back times are essential. 

Within the past ten years several large Hubble Space Telescope ({\it{HST}}) imaging surveys have been undertaken by various groups, each with their own goals and strategies, all utilizing the Advanced Camera for Survey's (ACS) high-resolution wide field camera (WFC) \citep{clampin2002}. The All-wavelength Extended Groth strip International Survey (AEGIS; \citealt{dav07}) is centered on the Extended Groth Strip (EGS) and is one of four fields targeted by the DEEP2 Galaxy Redshift Survey (\cite{davis03}, \cite{newman12}) for extensive spectroscopic follow up. The Cosmological Evolutionary Survey (COSMOS; \citealt{sco07}) was designed around the large single band {\it{HST}} survey with extensive follow-up spectroscopy from the zCOSMOS redshift survey \citep{lil09}. A major aim of the DEEP2, AEGIS and COSMOS surveys is to study galaxy evolution in the context of large scale structure. The Great Observatories Origins Deep Survey (GOODS) (\cite{dic03},\cite{giava04}) was designed to be one of the deepest {\it{HST}} imaging campaigns to date; with its small area but deep imaging, it was designed to probe galaxy evolution down to the faintest galaxies detectable. The Galaxy Evolution from Morphology and SEDs (GEMS) survey \citep{cal08} was designed to study galaxy evolution using multi-wavelength data to construct Spectral Energy Distributions (SEDs) and measure morphologies. 

The Advanced Camera for Surveys General Catalog (ACS-GC) unifies the largest {\it{HST}} ACS imaging surveys into a single, homogeneously analyzed data set. We used the Galaxy Analysis over Large Areas: Parameter Assesment by Galfitting Objects from SExtractor (G\textsc{alapagos}) code \citep{barden11}, which incorporates both G\textsc{alfit} \citep{pen02} and SE\textsc{xtractor} \citep{ber96} to construct photometric and morphological catalogs derived from the {\it{HST}} ACS imaging. We provide additional derived data products (e.g, color images, atlas images, G\textsc{alfit} residual images, and ACS FITS image cutouts) for every source in the catalog. We also provide redshifts collated from the various redshift surveys which accompany the imaging data for a large fraction of the sources. The main goal of the ACS-GC data set is to provide a large statistical sample of galaxies with reliable structural and distance measurements (for a subsample) to probe the evolution of galaxies over a wide range of look-back times. This data set can be utilized for various purposes, for example, these data have been used by \cite{geo09} to study the host galaxy morphologies of X-ray selected Active Galactic Nuclei (AGN) in the AEGIS, GOODS-S, and GEMS surveys. \cite{com09} reports the serendipitous  discovery of a dual AGN in the COSMOS field. \cite{pie10} study the effects an AGN has on host galaxy colours and morphological measurements. \cite{griffith10} study the morphological distributions of AGN selected using X-ray, radio, and IR imaging from the COSMOS survey. \cite{masters11} study the morphology of galaxies in the Baryon Oscillation Spectroscopic Survey. \cite{cooper11} study the impact of environment on the size evolution of massive early-type galaxies at intermediate redshift. \cite{holden11} study the evolution in the intrinsic shape distribution of early-type galaxies from $z\sim1$ to $z\sim0$. \cite{Welikala2012} study color gradients in galaxies out to $z\sim3$.

There are a handful of standard galaxy properties that are commonly quantified, such as: apparent magnitude, color, morphology/shape, redshift/distance, size, velocity dispersion, and metallicity. These can all be used to gain insight into the formation history and evolution of galaxies. Understanding how these properties change and evolve with redshift/time is integral in our construction of galaxy evolutionary models and scenarios. The combination of high-resolution, deep optical imaging and redshift measurements along with the structural parameters provided by the ACS-GC make it a powerful data set which can be used to study the evolution of galaxy structures over cosmic times. In $\S$2 we describe the imaging and redshift surveys used to construct the ACS-GC. We describe the redshift completeness and reliability in \S3. We give a brief description of the quantitative analysis in $\S$4. In $\S$5 we describe properties of the ACS-GC catalog, including the naming conventions and auxillary data products. We summarize this work in $\S$6. All magnitudes are given in the AB magnitude system.

\section{The Redshift and Imaging Data}

In this section we describe the {\it{HST}} ACS imaging used to construct the ACS-GC data set and give basic descriptions of the imaging properties. We also summarize the available redshifts acquired from the various surveys, both spectroscopic and photometric. We summarize the ACS imaging data in Table 1, giving central coordinates for the surveys, survey size, filters and pixel scales. Table 3 summarizes basic catalog statistics, giving number counts in the respective ACS filters as well as total number of spectroscopic (split by quality) and photometric redshifts.

\begin{table*}[h!tb!]
\begin{center}
\caption{ACS-GC survey fields}
\end{center}
\begin{center}
\begin{tabular}{ccccccc}\hline\hline
Survey&RA& DEC&area&Filters&pixel scale\\\hline
&(J2000)&(J2000)&(deg$^2$)&&$''$/pix\\\hline
AEGIS&$14\!:\!17\!:\!00\!$&$+52\!:\!30\!:\!00\!$&0.197&F606W \& F814W & 0.03\\
GOODS-N&$12\!:\!36\!:\!55\!$&$+62\!:\!14\!:\!15\!$&0.07&F606W \& F775W & 0.03\\
COSMOS&$10\!:\!00\!:\!28\!$&$+02\!:\!12\!:\!21\!$&1.8&F814W & 0.05\\
GEMS&$03\!:\!32\!:\!25\!$&$-27\!:\!48\!:\!50\!$&0.21&F606W \& F850LP & 0.03\\
GOODS-S&$03\!:\!32\!:\!30\!$&$-27\!:\!48\!:\!20$&0.07&F606W \& F850LP & 0.03

\\\hline
\end{tabular}
\begin{center}
\end{center}
\end{center}
\end{table*}

\subsection{The AEGIS Survey}

\subsubsection{Imaging}

The All-Wavelength Extended Groth Strip International Survey \citep{dav07} is a large collaborative effort designed to provide
one of the largest and deepest panchromatic data sets currently available. The region studied is centered on the Extended Groth Strip ($\alpha = 14^{\mathrm{h}}17^{\mathrm{m}}$, $\delta = 52^{\mathrm{o}}30'$), a region with deep observations covering all major wavebands from X-ray to radio. The {\it{HST}} ACS imaging in the EGS field is comprised of 63 pointings using both the F606W and F814W filters, with exposure times of 2260 and 2100 seconds, respectively, per pointing. The imaging covers a total area of $\sim$ 710 arcmin$^2$. Our analysis is based on images produced by the STSDAS {\it{multidrizzle}} package \citep{koek2002}, and the final images have a pixel scale of $0.03''$ per pixel. For an extended object the 5$\sigma$ limiting magnitudes are F606W=26.2 (AB) and F814W=25.6 (AB).

\subsubsection{Redshifts}

For the AEGIS survey we provide a total of 5,765 spectroscopic redshifts of which 4,244 are high-quality redshifts ($z_q \ge 3$) from the DEEP2 galaxy redshift survey data release 3 (DR3;\citealt{dav07}). DEEP2 targets were selected for spectroscopy from the CFHT 12K $BRI$ imaging described in \cite{dav07}. Eligible DEEP2 targets have 18.5 $\le R \le 24.1$ and surface brightness $\mu_R = R + 2.5 \log A < 26.5$, where $A$ is the area of the aperture (in sq. arcseconds) used to measure the CFHT 12K $R$-band magnitude. The DEEP2 catalog provides a quality metric ($z_q$) ranging from 1 for the lowest quality to 4 for the highest quality redshifts. Two significant features must match the spectral templates for a secure redshift (quality $z_q \ge 3$); Note that a resolved [O II]$ \lambda 3727$ doublet is counted as two features. The median redshift for the sample is 0.74. Galaxies at $z > 1.4$ generally lack strong features in the DEEP2 spectral window; these objects comprise the bulk of the DEEP2 redshift failures. Ongoing spectroscopic efforts in the field as part of the DEEP3 Galaxy Redshift Survey (\cite{cooper2011}; \cite{cooper12}) will significantly increase the completeness within the {\it HST}/ACS footprint. 

We also provide 43,796 photometric redshifts as described in \citep{coupon09}. Comparing with galaxy spectroscopic redshifts, in the wide fields, they find a photometric redshift dispersion of $0.037- 0.039$ and an outlier rate of 3-4\% at $i'_{\mathrm{AB}} < 22.5$. Beyond $i'_{\mathrm{AB}} = 22.5$ the number of outliers rises from 5\% to 10\% at $i'_{\mathrm{AB}} < 23$ and $i'_{\mathrm{AB}} < 24$, respectively. The redshift range $0.2 < z \le 1.5$ is the most suitable since this redshift range is better constrained by the filters used. 

\subsection{The GOODS Survey}

\subsubsection{Imaging}

The GOODS survey \citep{dic03,giav04} was designed to be a deep multi-wavelength data set with which to study the formation and evolution of galaxies. The GOODS survey targeted two separate fields, the $Hubble$ Deep-Field North (HDF-N) (now referred to as GOODS-N) and the $Chandra$ Deep-Field South (CDF-S) (now referred to as GOODS-S). The {\it{HST}} ACS imaging was carried out in four broad, non-overlapping filters, F435W($B$), F606W($V$), F775W($i$) and F850LP($z$). While the F435W images were all acquired at the beginning of the survey, the F606W, F775W, and F850LP were carried out in 5 epochs. The mean exposure time at each epoch was 1050, 1050, and 2100 s in the F606W, F775W, and F850LP bands, respectively. The imaging comprises 17 {\it{HST}} pointings in GOODS-N and 15 in GOODS-S.  Our analysis is based on images produced by the STSDAS {\it{multidrizzle}} package, and the final images have a pixel scale of $0.03''$ per pixel. We restrict our analysis to the F606W and F775W imaging in GOODS-N and the F606W and F850LP imaging in GOODS-S. For GOODS-S we analyzed the F850LP filter in order to combine directly with the GEMS F850LP imaging. The ACS imaging covers a total area of  $\sim$ 320 arcmin$^2$ (e.g., 160  arcmin$^2$ per field). The 5$\sigma$ limiting magnitudes for an extended source are F606W = 25.7, F775W = 25.0. 

\subsubsection{Redshifts}

For the GOODS-N survey we provide 2854 spectroscopic redshifts from various sources, of which 1347 are high-quality redshifts ($z_q \ge 3$). To keep track and organize the different sources for spectroscopic redshifts, we provide a parameter called Z\_ORIGIN. For $z\_origin$ equal to GOODS-N-ALL, refer to \cite{wir04} and \cite{cow04}; for the remainder of the spectroscopic redshifts, refer to \cite{bar08}.
 
 We provide 6,278 photometric redshifts as described in \cite{bun09}. Compared to spectroscopic redshifts, the photometric redshift outliers (defined by $|z_{\mathrm{spec}} - z_{\mathrm{phot}}| > 1$) account for 4\% of the redshift estimates, with $\sigma_{|\Delta z|/(1+z_{\mathrm{spec}}}) \approx 0.1$ when outliers are excluded.

\subsection{The COSMOS Survey}

\subsubsection{Imaging}

The Cosmological Evolution Survey (COSMOS) \citep{sco07} was designed to thoroughly probe the evolution of galaxies, AGNs, and dark matter in the context of their environment and to sample the full dynamic range of large-scale structure from voids to very massive clusters. COSMOS  acquired the largest contiguous {\it{HST}} ACS imaging survey to date, covering $\sim$ 1.8 deg$^2$ in the F814W filter. The original {\it{HST}} imaging consisted of 590 pointings. We use the publicly available mosaics described in \cite{koe07}. The total mean exposure times for each pointing is 2028 seconds. Our analysis is based on images produced by the STSDAS {\it{multidrizzle}} package \citep{koek2002}, and the final images have a pixel scale of $0.05''$ per pixel. For galaxies with half-light radii of $0.25''$, $0.50''$, and $1.00''$, the completeness is 50\% at F814W $\simeq$ 26.0, 24.7, and 24.5, respectively.

\subsubsection{Redshifts}

For the COSMOS survey we provide 10,236 spectroscopic redshifts, of which 8,472 are reasonably secure redshifts (confidence class 3.x, 4.x, 1.5, 2.4, 2.5, 9.3, 9.5, 13.x, 14.x, 23.x and 24.x) from the zCOSMOS redshift survey \citep{lil09}. The primary zCOSMOS targets were selected for spectroscopy from the ``total'' F814W magnitudes and were required to be in the magnitude range $15.0 < \mathrm{F814W} < 22.5$. The quality metrics used for the zCOSMOS survey are described in depth in Table 1 of \cite{lil09}. It is worth noting that only $\sim$5.0\% of the reasonably secure redshifts are at $z \ge 1.0$; the majority of the spectroscopic redshifts are in the range of $0.2 < z < 1.0$. Approximately 88\% of the galaxies observed in zCOSMOS have a spectroscopic redshift that is secure at the 99\% level.

We provide 251,971 photometric redshifts from \cite{ilb09}. These highly accurate photometric redshifts are based on 30-band photometry, which span the wavelength range of UV to mid-IR. Using a sample of 4,148 galaxies from the zCOSMOS-bright survey, \cite{ilb09} recover a catastrophic failure rate $\eta$ = 0.7 \% and redshift accuracy of $\sigma_{|\Delta z|/(1+z\mathrm{spec})}=0.007$ for $i^+$ $<$ 22.5. Due to the magnitude limits probed by the zCOSMOS-bright survey, photometric redshift reliabilities for fainter magnitudes, $i^+ > 22.5$, where $i^+$ refers to the Subaru photometric system, were tested using 209 galaxies from the zCOSMOS-faint survey and 317 galaxies from the MIPS spectroscopic sample \citep{kartal10}. At high redshift $1.5 < z < 3.0$,  \cite{ilb09}  recover a catastrophic failure rate $\eta = 20.4$ \% with a redshift accuracy $\sigma_{|\Delta z|/(1+z_{\mathrm{spec}})}=0.053$ with a median magnitude $i^+_{median} = 24.0$. For $22.5 < i^+ < 24.0$ they measure a redshift accuracy of $\sigma_{|\Delta z|/(1+z_\mathrm{sspec})}=0.011$. These results are summarized in Table 3 of \cite{ilb09}.

\subsection{The GEMS Survey}

\subsubsection{Imaging}

GEMS is an 800 arcmin$^2$ survey using the {\it{HST}} ACS instrument in two bands (V606W and F850LP); (\citealt{rix04}, \citealt{caldwell08}). The field was chosen due to the rich set of observations at complementary wavelengths. GEMS is centered on the Extended {\it{Chandra}} Deep Field South ($\alpha = 03^{\mathrm{h}}32^{\mathrm{m}}$, $\delta = -27^{\mathrm{o}}48'$). The central $\sim$ 25 \% of the E-CDFS field has deep {\it{HST}} ACS imaging from the GOODS survey. The {\it{HST}} ACS imaging in the GEMS field is comprised of 63 pointings using both the F606W and F850LP filters, with exposure times of 2160 and 2286 
seconds per pointing, respectively. Our analysis is based on images produced by the STSDAS {\it{multidrizzle}} package. The final images have a pixel scale of $0.03''$ per pixel. For an extended object the 5$\sigma$ limiting magnitudes are F606W=25.7 (AB) and F850LP=24.2 (AB).

\begin{table*}[h!tb!]
\caption{GEMS + GOODS-S spectroscopic references}
\begin{center}
\begin{tabular}{cc}\hline\hline
\it{z\_origin}& Reference\\
\hline
VLT\_2008 & \cite{vanzella08}\\
VLT\_IMAG & \cite{Ravikumar07}\\
VLT\_LBGs & \cite{Vanzella09}\\
VIMOS\_08\_MR/LR & \cite{Popesso09}\\
GRISM\_HUDF & \cite{hathi08} and \cite{Rhoads09} \\
ePEARS\_HUDF & \cite{Straughn08} \\
ePEARS\_CDFS & \cite{Straughn09} \\
GRAPES\_HUDF & \cite{hathi09} and \cite{Pasquali06}\\
K20 &\cite{Mignoli05} \\
CXO-CDFS & \cite{Szokoly04} \\
VVDS & \cite{Lefevre04} \\
LCIRS & \cite{doherty05} \\ 
FW\_5 & \cite{norman02} \\
FW\_6 & \cite{croom01} \\
FW\_7 & \cite{vanderwal05} \\
FW\_8 & \cite{Cristiani00} \\
FW\_9 & \cite{Strolger04} \\
FW\_10 & \cite{Daddi04} \\
FW\_13 & \cite{wuyts09} \\
FW\_14 & \cite{Kriek08} \\
FW\_15 & \cite{Roche06} \\
FW\_16 & \cite{wuyts08}\\
ACES& \cite{cooper12}
\\\hline
\end{tabular}
\begin{center}
\end{center}
\end{center}
\end{table*}

\subsubsection{Redshifts}

For the GEMS + GOODS-S surveys we provide spectroscopic redshifts from various sources (Table 2) and provide a total of 6,955 spectroscopic redshifts, with 5,756 high-quality redshifts ($z_q \ge 3$). The quality of the redshifts range from 1 for the lowest quality to 4 for the highest quality redshifts. Refer to the catalog parameter Z\_ORIGIN for the origin of the spectroscopic redshift (see Table 2).

We provide 44,239 photometric redshifts from the COMBO-17 survey \citep{wolf08}. Using a high-quality subset of spectroscopic redshifts from \cite{Lefevre04}, they find the $\Delta z/(1+z_{\mathrm{s}})$ deviations to have an rms $\sim$ 0.008 at $R < 21$, increasing to 0.02 at $R < 23$, and 0.035 for $23.0 < R < 24.0$. Note, however, not much is known about the photometric redshift accuracy for normal galaxies at $ z > 1.2$. Refer to \cite{wolf04} and \cite{wolf08} for a full description of this data.

\section{Redshift Completeness and Reliability}

All extra-galactic surveys are fundamentally limited by the completeness in their spectroscopic and photometric redshifts. Referring to Table 3, we can see that the redshifts for each survey are dominated by the photomteric redshifts and these will dominate the completeness of the redshift survey. Thoughout this particular analysis we concentrate on the photometric redshift samples, focusing on the highest reliable photometric redshifts provided by each survey. In Figure 1 we plot the histograms of the photometric redshift errors provided by each survey. The AEGIS, GEMS, and GOODS-S are 1$\sigma$ and COSMOS is 3$\sigma$. We observe a peculiar bi-modal distribution for the GEMS and GOODS-S  distribution. The photometric redshift errors for GOODS-N are larger than for the other surveys. Users should exercise caution when using this sample.. To select reliable photometric redshifts from AEGIS, GEMS and GOODS-S, and COSMOS we require photoz\_err $\le 0.15\times(1+ photoz)$.

\begin{figure*}[h!tb]
\begin{center}
\begin{tabular}{c}
\includegraphics[scale=0.6,angle =90]{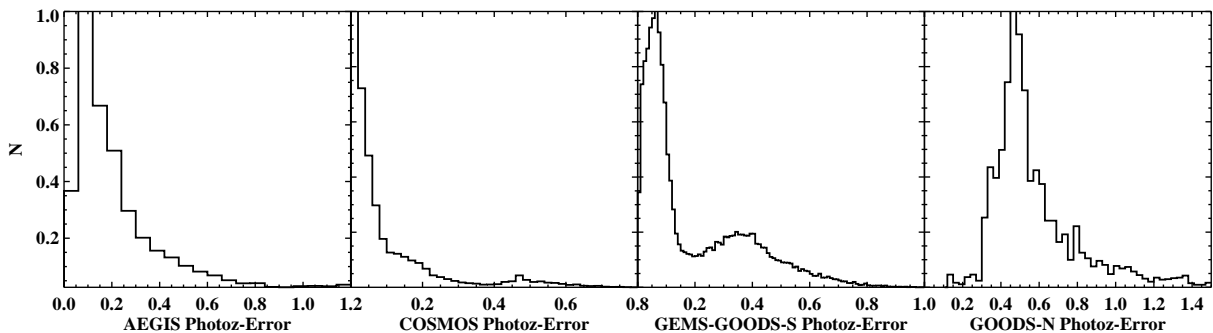}
\end{tabular}
\end{center}
\caption[CAPTION]{\label{fig:1} Histograms of the photometric redshift errors in the ACS-GC surveys.}
\end{figure*}

Another reliability test that can be done is to compare high quality spectroscopic redshifts to their photometric counterparts. We select all spectroscopic redshifts with $z_q \ge 3$, while the COSMOS high-quality redshifts are described in Section 2.3.2. In Figure 2 we plot the high quality spectroscopic redshift versus the photometric redshift. We observe EGS and COSMOS to have highly consistent results, while GEMS and GOODS-S seem to have larger uncertainties at $z > 1.0$. The GOODS-N sample appears to have the largest dispersions, and users should exercise caution when using this sample.

\begin{figure*}[h!tb]
\begin{center}
\begin{tabular}{c}
\includegraphics[scale=0.6,angle =90]{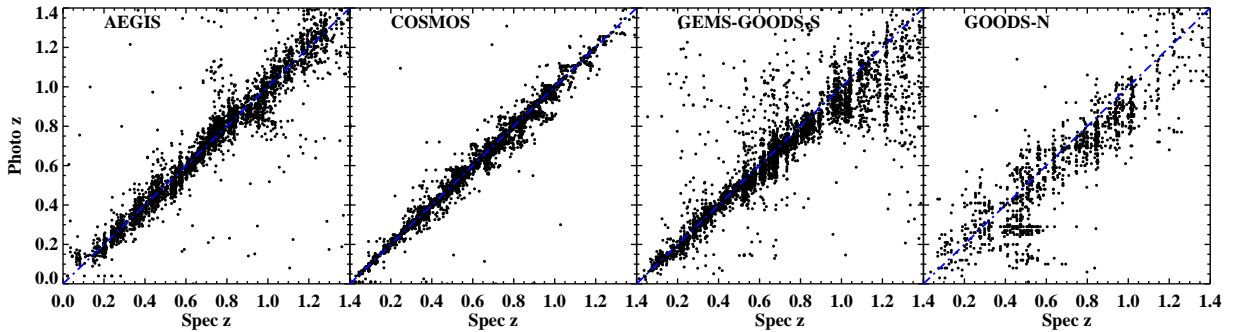}
\end{tabular}
\end{center}
\caption[CAPTION]{\label{fig:1} Spectroscopic redshift versus photometric redshift in the ACS-GC surveys.}
\end{figure*}

Having an unbiased estimation of the redshift completeness requires reliably removing compact sources and Low Surface Brightness (LSB) galaxies from the sample. 
This is performed by utilizing the method described in \S 5.5. Using  a sample of $normal$ extended galaxies with reliable photometric redshift estimates we can estimate the redshift completeness as a function of magnitude in the following  manner. The completeness for a given magnitude bin ($\Delta$ Mag 0.5) is given by

\begin{equation}
C(mag) = \frac{N(z)}{N(total)}
\end{equation}

\begin{figure*}[h!tb]
\begin{center}
\begin{tabular}{c}
\includegraphics[scale=0.7]{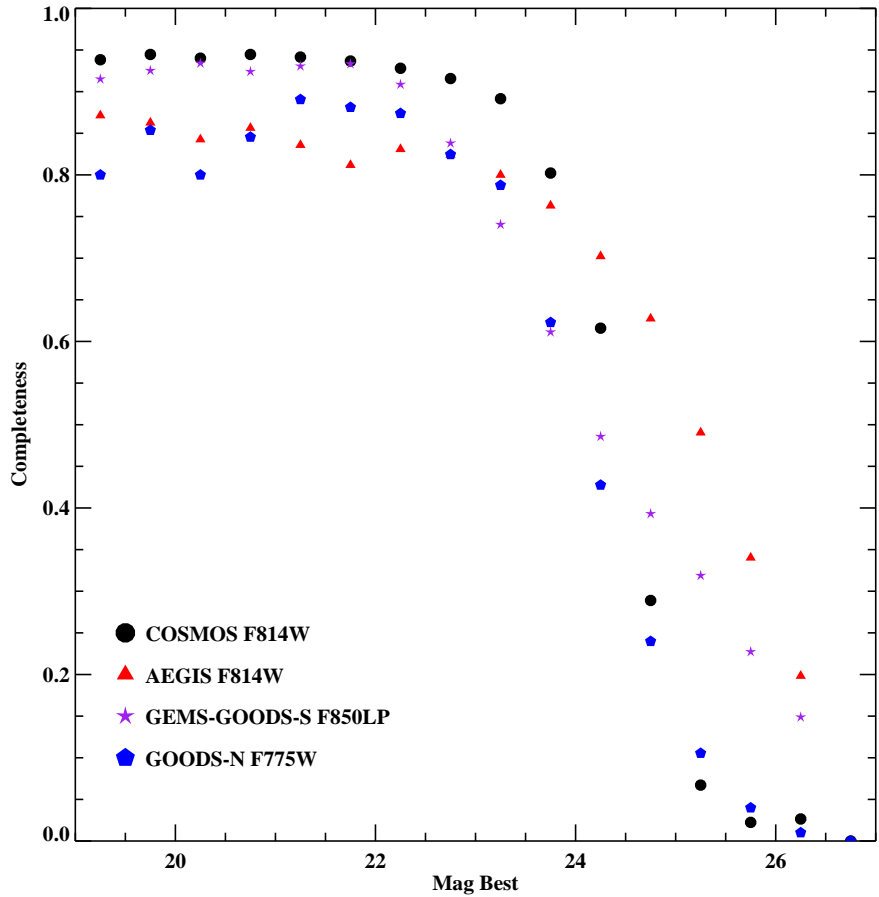}
\end{tabular}
\end{center}
\caption[CAPTION]{\label{fig:1} Mag Best versus completeness for all surveys in the ACS-GC. The selection of the photometric redshifts are described \S 3}
\end{figure*}

In Figure 3 we plot the photometric redshift completeness as a function of magnitude for all surveys. For AEGIS and COSMOS we plot F814W, for GEMS and GOODS-S we plot F850LP, and for GOODS-N we plot F775W. For COSMOS we can see that the sample is highly complete to F814W $<$ 23.5 and dropping to 75\% at F814W = 24.0. AEGIS is highly complete to F814W $<$ 23.0 and drops to 70\% at F814 = 24.0. GEMS and GOODS-S are a bit shallower than AEGIS and COSMOS, being 75\% complete at F850LP = 23.5. For GOODS-N we did not apply any reliability criteria and compute the completeness with all available measurements. We can see that GOODS-N is 75\% complete at F775W = 23.5.

\section{Galaxy Photometry and Quantitative Morphology}

In order to combine and analyze this extremely large imaging data set we adopted an automated fitting method called G\textsc{alapagos}. G\textsc{alapagos} was written in the IDL language to analyze large ACS imaging data sets through the G\textsc{alfit} code \citep{barden11}. The code was tested and compared to the Galaxy Image 2D (GIM2D) \citep{simard02} code by \cite{hau07} using the F850LP GEMS ACS imaging. They conclude that G\textsc{alfit} is more robust in crowded fields since it does simultaneous fitting of nearby galaxies, a capability not available with GIM2D. G\textsc{alfit} and GIM2D use different convergence methods and \citep{hau07} found G\textsc{alfit} operated faster than GIM2D in analyzing these large imaging surveys. G\textsc{alapagos} is structured into four program blocks: SExtraction, postage stamp cutting, sky estimation and G\textsc{alfit}, and catalog creation. The code is controlled mainly through a setup script and a file location list. Refer to \cite{hau07} for a detailed description of G\textsc{alapagos}. We next give a brief description of our SE\textsc{xtractor} and G\textsc{alfit} setup.

\subsection{SE\textsc{xtractor}}

We use SE\textsc{xtractor} \citep{ber96} to create the photometric catalogs used as initial inputs given to G\textsc{alfit}. SE\textsc{xtractor} detects, deblends, measures and classifies objects, giving estimates of magnitude, size, axis ratio ($b/a$), position angles, and a star-galaxy classification. The GEMS team found that no single SE\textsc{xtractor} setup satisfactorily detected and deblended both bright, well-resolved galaxies and faint galaxies near the detection limit. Accordingly, the best setup found by GEMS was to run SE\textsc{xtractor} twice: once to detect bright objects without splitting them up (what is called the `cold' mode) and once to detect faint objects (`hot' mode). The two modes are then combined to give one single catalog containing all objects. The procedure is described in detail in \cite{rix04} and \cite{cal08}. We use the final combined catalog to provide G\textsc{alfit} with initial input parameters. 

\subsection{G\textsc{alfit}}

G\textsc{alfit} is designed to measure structural parameters from galaxy images. We model each source in the catalog with a single S\'ersic profile as well as a model for the sky (which we keep fixed during the fit). The S\'ersic profile (1968) is defined as
\begin{equation}
\Sigma (r) = \Sigma_e e^{-k[(r/r_e)^{1/n} - 1]}
\end{equation}

\noindent
where $r_e$ is the effective radius of the galaxy, $\Sigma_e$ is the
surface brightness at $r_e$, $n$ is the S\'ersic index, and $k$ is
coupled to $n$ such that half of the total flux is always within
$r_e$. Before evaluating its fit to the data, G\textsc{alfit} convolves the 2-D image with
a Point Spread Function (PSF), derived empirically from a high S/N star, with a single PSF used for each band and survey. G\textsc{alfit} then uses a Levenberg-Marquardt Algorithm for
$\chi ^2_\nu$ minimization. The S\'ersic profile has seven free parameters: $x-center$, $y-center$, $position$ $angle$,  {\it{S\'ersic}} $index$, {\it{half-light}} $radius$, $axis$ $ratio$, and $magnitude$. G\textsc{alfit} requires a setup script, which is created by G\textsc{alapagos}, which has initial guesses for many of the parameters. In particular, using SExtractor parameters, starting magnitudes were given by MAG\_BEST, sizes were derived from the FLUX\_RADIUS using the formula $r_e = 0.162R^{1.87}_{\rm flux}$ , where $R_{\rm flux}$ is FLUX\_RADIUS. This formula was determined empirically using simulations. The axis ratio b/a and the position angle were derived by taking the SExtractor parameters ELLIPTICITY and THETA\_IMAGE, respectively. Furthermore, the position of each object within its postage stamp was required as an input parameter for G\textsc{alfit}, which was directly given by the process of cutting the postage stamps (the object is centered within its postage stamp). See \cite{hau07} for a more detailed description of this process. Our initial input for the S\'ersic index was 2.5. G\textsc{alfit} produces a summary of the fit parameters as well as a FITS image block which includes the original image, the model image, and the residual image (original -- model).

\begin{table*}[h!tb!]
\caption{Catalog Statistics}
\begin{center}
\begin{tabular}{ccccccc}\hline\hline
Survey&Objno&Filter&Ntot&specz&specz($z_q \ge 3$)&photoz\\\hline
AEGIS&1xxxxxxx&F606W&65,301&5,691&4,244&41,982\\
&&F814W&55,808&5,691&4,244&37,294\\
&&F606W+F814W&50,967&5,691&4,244&35,480\\\hline
COSMOS&2xxxxxxx&F814W&304,688&10,236&8,472\footnotemark[2]&251,971\\\hline
GOODS-N&5xxxxxxx&F606W&23,071&2,793&1,332&6,051\\
&&F775W&17,592&2,832&1,343&6,128\\
&&F606W+F775W&16,438&2,771&1,328&5,901\\\hline
GEMS-GOODS-S&9xxxxxxx&F606W&63,321&6,792&5,639&42,942\\
&&F850LP&54,613&6,781&5,694&37,613\\
&&F606W+F850LP&47,488&6,618&5,577&36,316\\\hline
\end{tabular}
\end{center}
\end{table*}

\section{The Catalog}
For each ACS survey we combined the SE\textsc{xtractor}, G\textsc{alfit}, and redshift catalogs to produce a single combined catalog. We then combined all of the surveys to produce the single, uniformly constructed ACS-GC catalog\footnote[1]{www.ugastro.berkeley.edu/~rgriffit/Morphologies/}. This catalog has 97 parameters, in order to provide a comprehensive list of galaxy properties. Table 5 presents a description of the parameters.  We use an NGC-style numbering scheme, refer to \S5.1 and Table 4. The naming convention is similar to the DEEP2 redshift survey. We also unite the photometry and structural measurements for the different surveys in a consistent manner by appending \_HI and \_LOW to parameters which were measured in the individual ACS filters, where  \_LOW refers to the F606W filter while \_HI is F850LP for GEMS and GOODS-S, F775W for GOODS-N and F814W for COSMOS and AEGIS. The catalog parameter IMAGING gives the origin of the ACS imaging used to measure the parameters of interest, and is useful for separating GEMS and GOODS-S.

\footnotetext[2]{See section 2.3.2 for a description of this sample}

\noindent
Table 3 gives basic catalog statistics, e.g., object numbers, filters, total number of sources identified in each filter (Ntot), total number of spectroscopic redshifts, total number of high quality spectroscopic redshifts ($z_q \ge 3$) and total number of photometric redshifts in each filter.

\subsection{Object Identification}

Our object identification scheme has been adopted from the DEEP2 survey, which uses an 8 digit number to identify each source in the catalog. The convention is motivated by the fact that each input survey uses its own naming convention. Combining these surveys into one homogeneous data set required creating a single, uniform naming convention across all surveys. Table 4 gives a description of the object numbers and naming convention for the individual surveys.

\begin{table*}[h!tb!]
\caption{Object Numbers}
\begin{center}
\begin{tabular}{ccl}\hline\hline
Survey&Objno&Description\\\hline
AEGIS&100xxxxx& F814W \& F606W Detection in ACS-GC but not DEEP2\\
&101xxxxx&F814W detection only in ACS-GC but not DEEP2\\
&102xxxxx&F606W detection only in ACS-GC but not DEEP2\\
&1(1/2/3/4)0xxxxx& F814W \& F606W Detection in ACS-GC and DEEP2\\\hline
COSMOS&20xxxxxx&F814W detection\\\hline
GOODS-N&500xxxxx&F775W and F606W detection\\
&501xxxxx&F775W detection only\\
&502xxxxx&F606W detection only\\\hline
GEMS + GOODS-S&900xxxxx&F606W and F850LP detection\\
&901xxxxx&F850LP detection only\\
&902xxxxx&F606W detection only\\\hline

\end{tabular}
\end{center}
\end{table*}

We also supply the ``SURVEY\_ID"  parameter in the ACS-GC catalog, which is the ID number used by the original survey. This allows users to easily and rapidly match the ACS-GC catalog, rather than having to cross-correlate catalogs using positions. This parameter is given, where available, for the AEGIS, COSMOS, and GOODS-N surveys. We do not provide this for GEMS and GOODS-S due to naming convention used by these teams, which was using the source position, ra and dec as the source ID.

To improve computational efficiency, some of the fields were divided into tiles with a small overlap between them, to ensure no objects were lost. Because of this, some objects appear more than once when merging catalogs of sources in the ACS-GC. These duplications were removed by coordinate matching and visual inspection. Nevertheless, some repeated objects may still exist in the final catalogs, but the number should be very small and will be completely dominated by objects close to the faint detection limit.

\subsection{Flags}

We use a very simple method to distinguish whether a source has a good fit (FLAG =0) or an unreliable fit (FLAG = 1). We use the G\textsc{alfit} uncertainties for both the half-light radius and the S\'ersic index $n$, and we use CLASS\_STAR to separate extended sources from compact sources. Our good fits (FLAG = 0) require $\sigma (n) \le  0.15*n$,  $\sigma (r_e) \le 0.15*r_e$, and CLASS\_STAR $\le 0.8$. The additional requirement given by CLASS\_STAR assigns unreliable results for both stellar like and compact objects. As the source size becomes comparable to the PSF size the results become increasingly unreliable. Since this is a very simple cut using few uncertainty parameters, the users of this data set are advised to use as many uncertainty parameters ($\chi^2_\nu$, surface brightness, magnitude limited samples, etc), to define high-quality samples for their investigations.

\subsection{Reliability and Measurement Errors}

Structural parameter errors quoted in the ACS-GC come directly from the G\textsc{alfit} fitting results. It is worth noting that \cite{hau07} found that G\textsc{alfit} substantially underestimated the true fit uncertainties, indicating that the dominant contribution to the fitting uncertainties is not shot and read noise, but instead contamination from neighbors, structure in the sky, correlated pixels, profile mismatch, etc. They also find that the reliability of the fitting results was dependent on the galaxy type measured. For galaxies with exponential profiles ($n =1.00$) and brighter than the sky's surface brightness, they found no significant mean offset between the input and recovered parameters. For galaxies exhibiting a de Vaucouleurs profile ($n =4.00$), they find that G\textsc{alfit} recovers parameters that are significantly less accurate than the $n =1.00$ galaxies. This behavior is attributed to two factors. First, spheroidal profiles are in principle harder to fit due to the importance of the outskirts of the light profile, thus requiring a careful and accurate measurement of the sky background to be used in order to return a reliable fit. Second, due to the large amount of light in the faint wings of the galaxies, neighboring objects have a much bigger influence on the fit of the galaxy of interest. 

\begin{figure*}[h!tb]
\begin{center}
\begin{tabular}{c}
\includegraphics[scale=0.4]{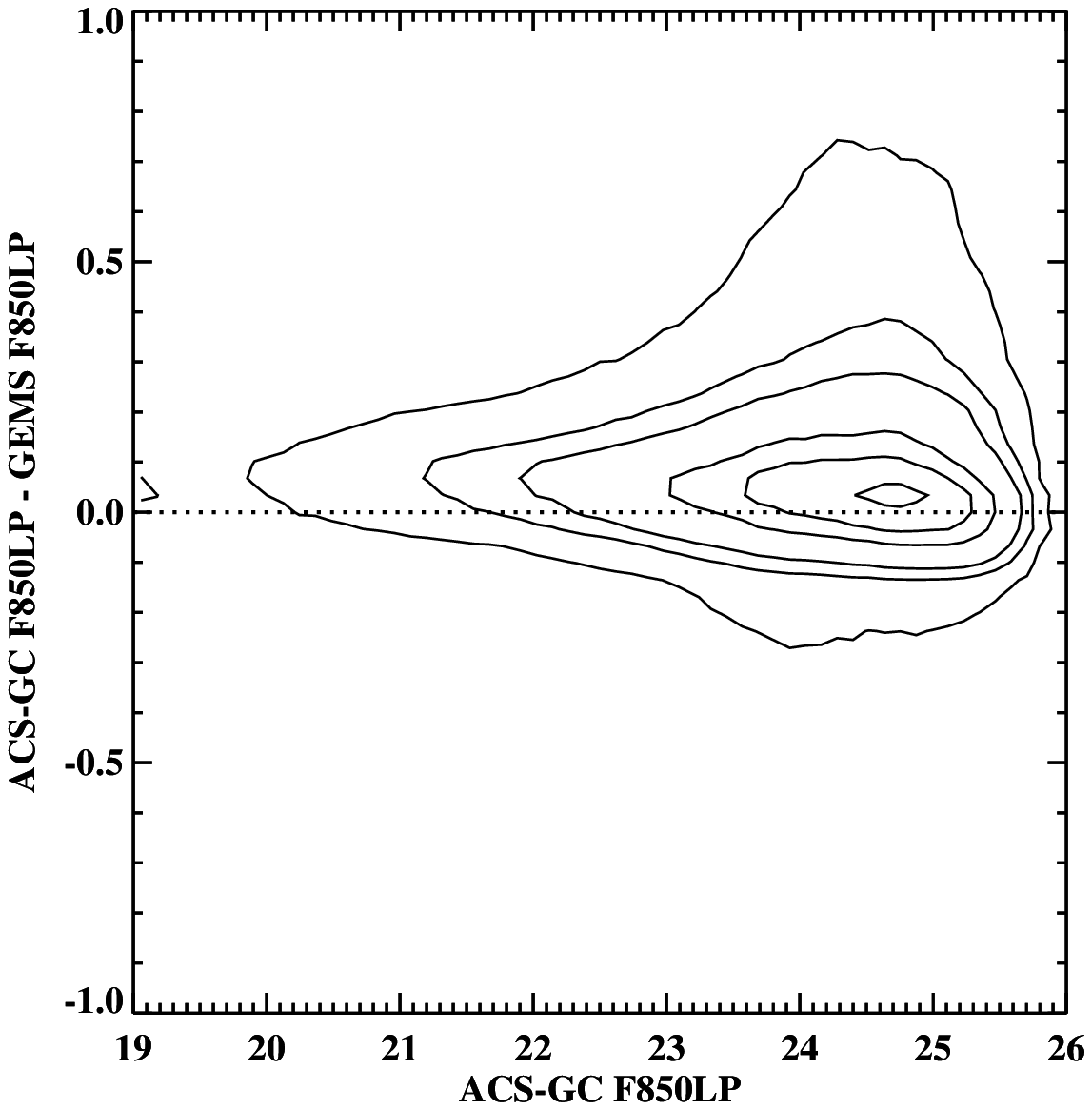}
\includegraphics[scale=0.4]{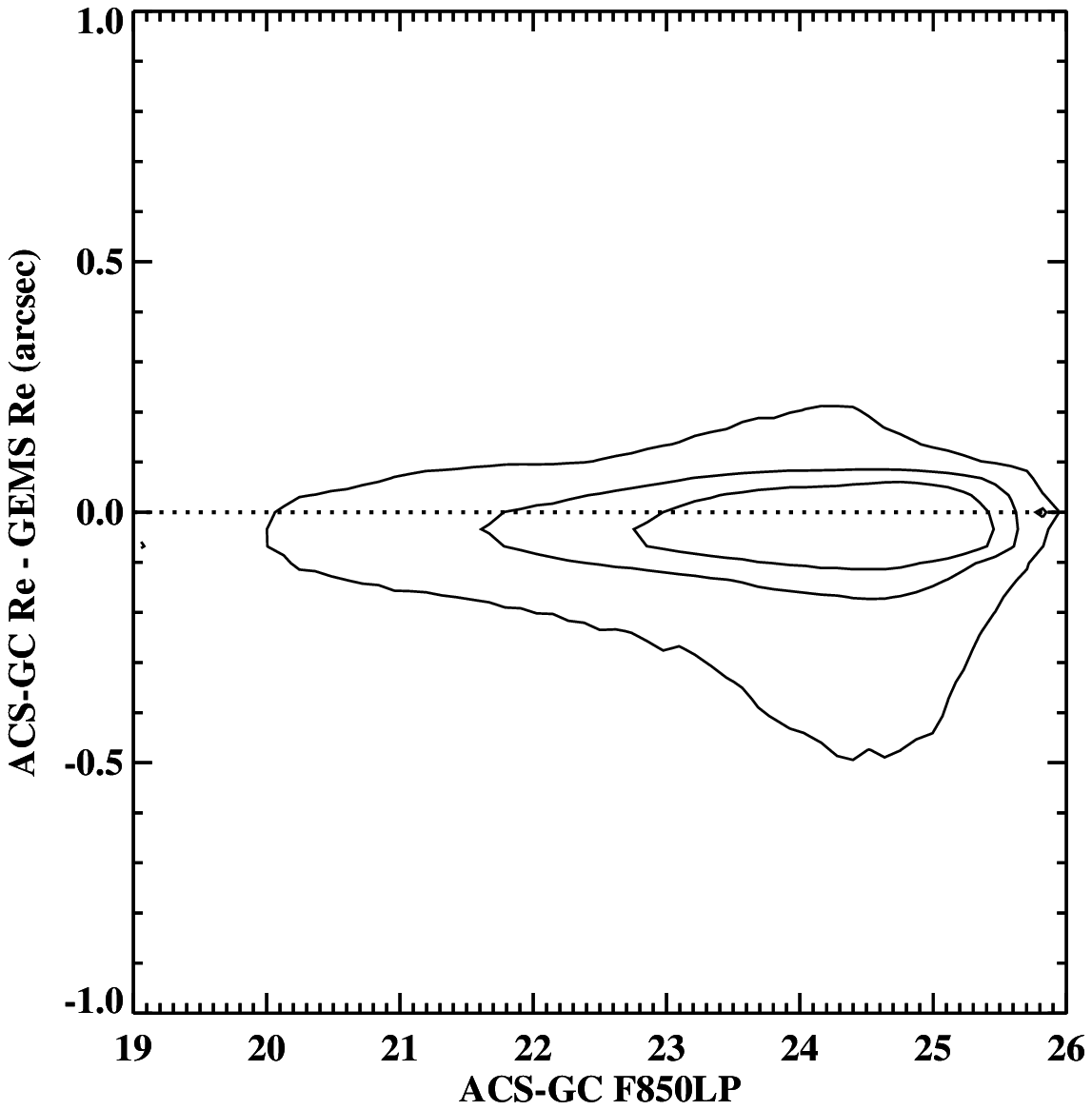}
\includegraphics[scale=0.4]{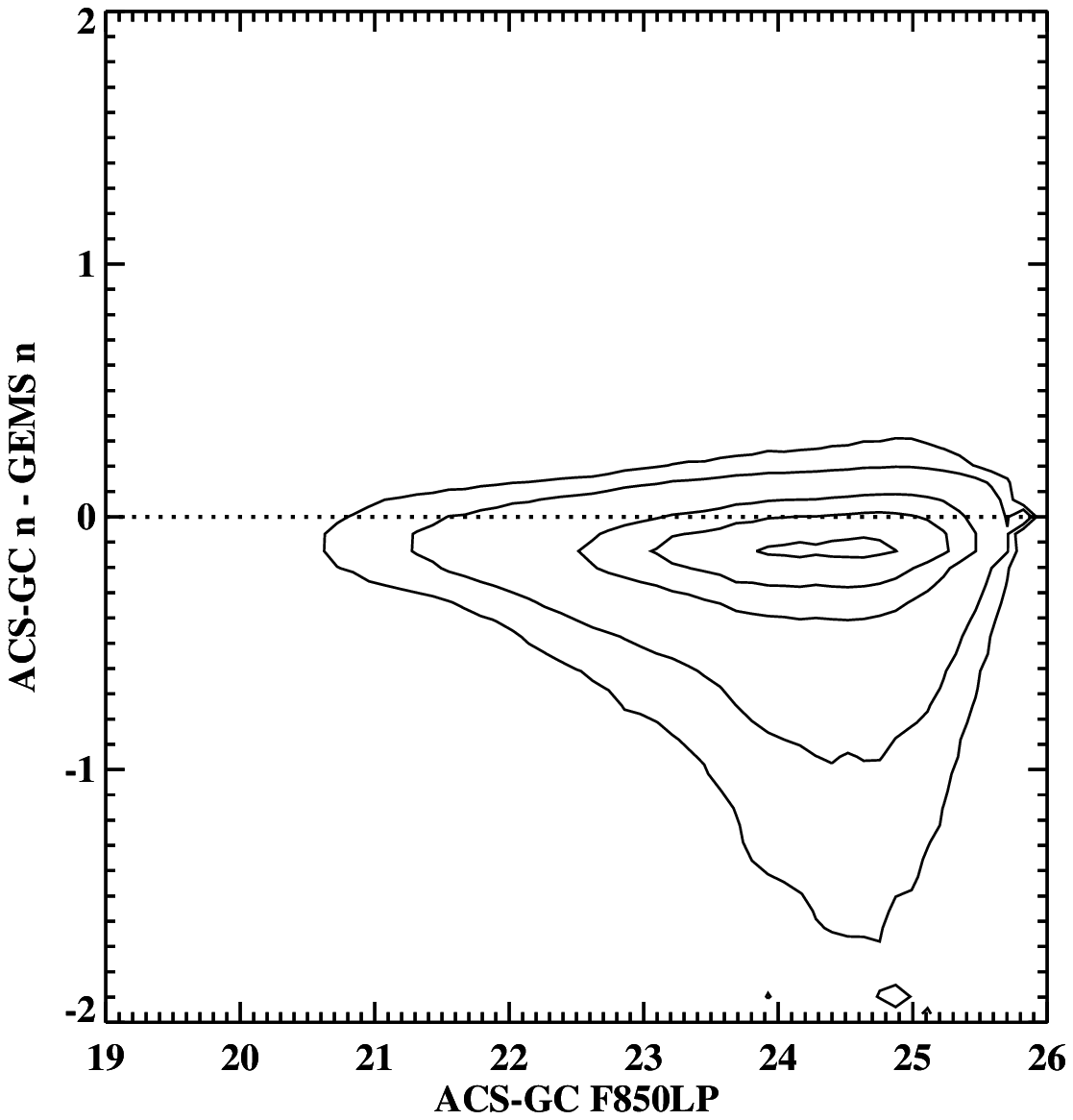}
\end{tabular}
\end{center}
\caption[CAPTION]{\label{fig:1} Left: $\Delta$ Mag G\textsc{alfit} versus ACS-GC F850LP. Center: $\Delta r_e$ versus ACS-GC F850LP. Right: $\Delta n$ versus ACS-GC F850LP }
\end{figure*}

As a sanity check on the structural parameters of the ACS-GC we compare the results from the ACS-GC GEMS F850LP imaging to the results obtained in \cite{hau07}. Figure 4 shows the comparison between these two analyses. Left plot gives $\Delta$F850LP versus F850LP, center plot gives $\Delta r_e$ versus F850LP, and right plot gives $\Delta n$ versus F850LP. As expected, we observe a clear systematic trend in the differences of the recovered parameters as a function of magnitude. The recovered parameters are highly consistent to F850LP $\le$ 24.0. Sources with F850LP $\ge$ 24.0 show larger systematic differences, especially the magnitudes and S\'ersic index measurements. These results show that for galaxies above the sky's surface brightness the recovered parameters are generally reliable, but for fainter galaxies users should apply caution when using the derived parameters.

\subsection{Auxiliary Parameters}

In addition to parameters measured with the ACS images we provide a few additional useful parameters. We include the CFHTLS ($u,g,r,i,z$) photometry (COSMOS and AEGIS) \citep{gwy08}. We also provide $BRI$ magnitudes for both COSMOS \citep{cap07} and AEGIS \citep{dav07}. The parameter Ntot, which was derived during the catalog creation process gives the number of sources which were simultaniously fit with G\textsc{alfit} while fitting the primary source. This could be used to investigate line-of-sight over-densities in the ACS imaging. We also supply the surface brightness, defined as

\begin{equation}
\mu = {\rm mag} + 2.5\cdot(\log(2\cdot b/a\cdot\pi\cdot(r_e)^2)
\end{equation}

\noindent
where mag is given by Mag Best, $b/a$ is the axis ratio, and $r_e$ is given in arc-seconds. These parameters can be useful in the investigation of detailed galaxy properties and selecting complete and reliable samples, see \S 5.5

\subsection{Compact and Extended Sources in the ACS-GC}

\begin{figure*}[h!tb]
\begin{center}
\begin{tabular}{c}
\includegraphics[scale=0.6]{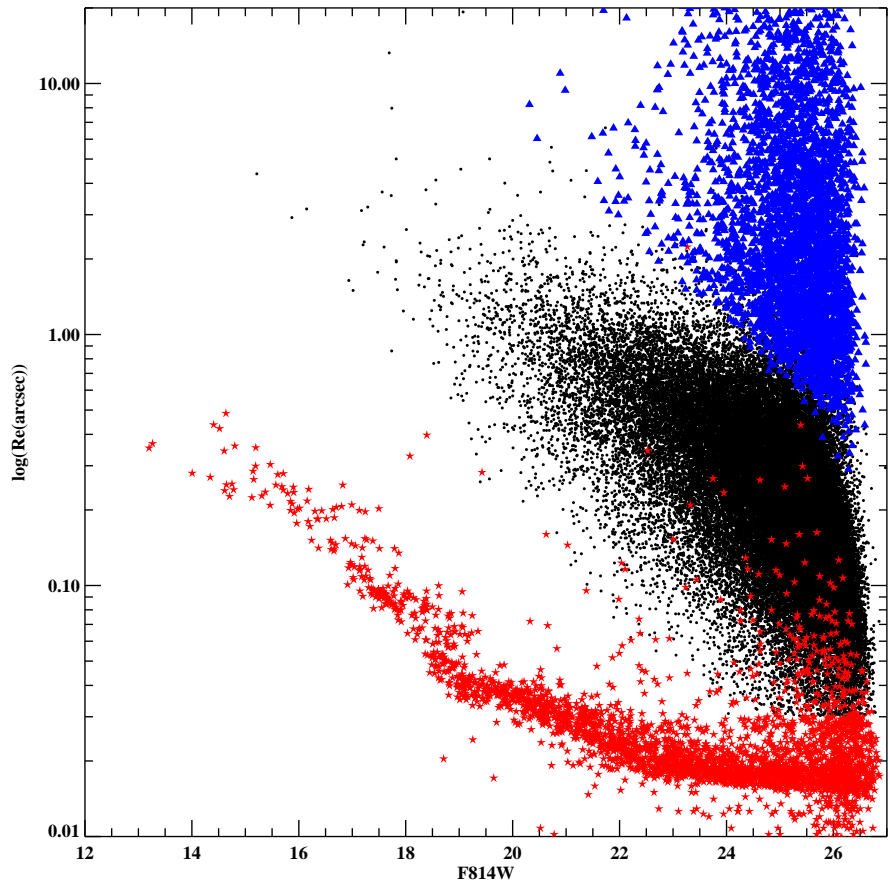}
\end{tabular}
\end{center}
\caption[CAPTION]{\label{fig:1} F814W versus $\log$ $r_e''$ for all galaxies in the AEGIS survey with detections in the F814W filter. Red stars are compact sources, black dots are extended sources, and blue triangles are LSB's.}
\end{figure*}

It has been known that the CLASS\_STAR parameter returned by SE\textsc{xtractor} is problematic in reliably distinguishing compact stellar-like sources and extended sources in imaging surveys. By defining compact objects as those having $\mu \le 18$ or ($\mu \ge 18$ and $r_e \le 0.03''$) we easily circumvent this issue. We demonstrate the reliability of this definition in Figure 5, where we plot all sources in the AEGIS survey having a F814W detection, with the x-axis representing the F814W magnitude and the y-axis the half-light radius $r_e$ given in arcseconds. Red stars represent compact sources (by our definition) and black circles represent extended sources. There is however another class of galaxies which have been notorious for producing unreliable results, these are the Low Surface Brightness (LSB) galaxies and tend to populate the top right hand corner of the magnitude-size diagram. These can easily be removed by requiring the extended galaxies to have $\mu < 26.0$. The extended galaxies with $\mu > 26.0$ are represented by the blue triangles in Figure 5 and are considered to be LSB's. We can see that these definitions do an excellent job in distinguishing between these three populations. Similar cuts can be applied to all the ACS-GC surveys to separate compact sources from extended sources and LSB galaxies.

\subsection{G\textsc{alfit} Residiual Maps, Color Images, and The Galaxy Atlas}

We provide high-resolution ACS pseudo-color images for the GEMS, AEGIS, and GOODS surveys, from which two-band imaging was available. These RGB images were made using the F814W and F606W images for the AEGIS data, the F850LP and F606W images for GEMS + GOODS-S, and the F775W and F606W images for GOODS-N. For example, the AEGIS color images were made using the following convention: the red channel was assigned to the F814W image, the blue channel was assigned to the F606W image and the green channel was assigned to (F814W+F606W)/2. These individual images were then converted into color images using the  IDL routine djs\_rgb\_make.pro (David Schlegel, personal communication). The COSMOS survey only has a single ACS band (F814W), thus making it impossible to derive ACS high-resolution color images. However, pseudocolor images in the COSMOS field were constructed by P.  Capak using the ACS F814W data as an illumination map and the Subaru  $B_J$, $r^+$, and $i^+$ images as a color map. To achieve this, each Subaru image was divided by the average of the three Subaru images and then multiplied by the ACS F814W image. This preserves the flux ratio between images while replacing the overall illumination pattern with the F814W data. Each image was then divided by $\lambda^{2}$ to enhance the color difference between star-forming and passive galaxies. The processed  $B_J$, $r^+$, and $i^+$ images were then assigned to the blue, green, and red channels, respectively. The resulting images have the high spatial resolution of the ACS imaging but color gradients at ground-based resolution. For every source in the ACS-GC catalog we provide a high-resolution color image as well as the original ACS FITS images used to make the color images. For COSMOS we also provide the Subaru images used to make the color images.

\begin{figure*}[h!tb]
\begin{center}
\begin{tabular}{c}
\includegraphics[scale=0.80,angle=90]{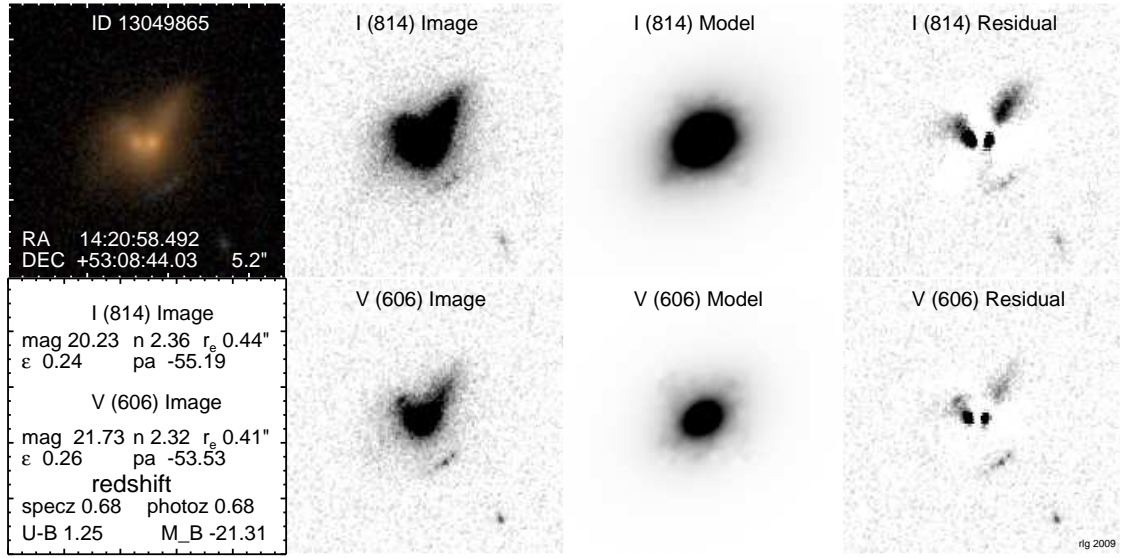}
\end{tabular}
\end{center}
\caption[CAPTION]{\label{fig:1} An example ACS-GC atlas image, as described in $\S$5.6}
\end{figure*}

For every source fitted by G\textsc{alfit}, G\textsc{alfit} returns a FITS image block which contains four extensions. Extension = 0 is blank, extension = 1 is the original ACS image, extension = 2 is the G\textsc{alfit} model image, and extension = 3 is the G\textsc{alfit} residual image (model $-$ original). These residual images are useful for many applications. For example, they can be used to identify rare classes of galaxies, such as gravitational lenses, ring galaxies, dual AGN \citep{com09}, and mergers. The residual maps also allow a visual confirmation of the quality of the fit. For every source in the ACS-GC catalog we provide this G\textsc{alfit} image block. for the GEMS, AEGIS, and GOODS survey this is generally two files, one for each band. The COSMOS single-band imaging produces only one of these files.

We combine this secondary imaging data as well as key strctural parameters into a single file for each source, the atlas image. Figure 6, which shows the atlas image of ACS-GC 13049865 (AEGIS) as an example, provides the ACS color image (top left) and the G\textsc{alfit} image blocks with the redder band in the top row and the bluer band, when available, as the lower row. The color panel provides the object number, RA, DEC, and the field of view in arc-seconds. The bottom left panel gives key parameters for the source, including the magnitude (Mag Best), the S\'ersic index ($n$), the half-light radius ($r_e$), the ellipticity ($\epsilon$), and the position angle (PA) for each band analyzed. We also give the spectroscopic and photometric redshifts (when available) and the $U-B$ rest-frame color and absolute $B$-band magnitude. 

\section{Summary}

In this work, we have measured photometric and structural parameters for $\sim$ half a million galaxies contained within the largest {\it HST} ACS extragalactic imaging surveys obtained to date.  These surveys have not been analyzed in a consistent manner previously.  The unified analysis presented here opens the possibility for scientific investigations that rely on these multiple fields being analyzed in a consistent manner.  We publicly release the ACS-GC catalog which includes 97 parameters for 469,501 astronomical sources, as well as secondary science products such as color images, G\textsc{alfit} images, atlas images, and FITS images (with WCS). Additional data products are expected in the near future from the Galaxy Zoo\footnote{www.galaxyzoo.org} project, who plan to visually classify a large fraction of the ACS-GC color images. The ultimate goal of the ACS-GC galaxy morphology data set is to provide a statistically significant sample of galaxies which can be used to investigate detailed galaxy properties as well as to understand how galaxy structures evolve over cosmic times. 

We gratefully acknowledge the principal investigators responsible for the ACS imaging utilized by the ACS-GC. We also acknowledge the GEMS team for creating and sharing the G\textsc{alapagos} code, without which none of this would have been possible. Work on this paper was carried out at the Jet Propulsion Laboratory and Infrared Processing and Analysis Center (IPAC), California Institute of Technology, under contract with NASA and WISE. J.M.C. is supported by an NSF Astronomy and Astrophysics Postdoctoral Fellowship under award AST-1102525. We thank the anonymous referee for his/her useful suggestions which have improved this manuscript.

Based on (GO-10134, GO-09822, GO-09425.01, GO-09583.01, GO-9500) program observations with the NASA/ESA Hubble Space Telescope, obtained at
the Space Telescope Science Institute, which is operated by the Association of Universities for Research in
Astronomy, Inc., under NASA contract NAS 5-26555. Based on observations obtained with MegaPrime/MegaCam, a joint project of CFHT and CEA/DAPNIA, at the Canada-France-Hawaii Telescope (CFHT) which is operated by the National Research Council (NRC) of Canada, the Institut National des Science de l'Univers of the Centre National de la Recherche Scientifique (CNRS) of France, and the University of Hawaii. This work is based in part on data products produced at TERAPIX and the Canadian Astronomy Data Centre as part of the Canada-France-Hawaii Telescope Legacy Survey, a collaborative project of NRC and CNRS. Funding for the DEEP2 Galaxy Redshift Survey has been provided in part by NSF grant AST00-71048 and NASA LTSA grant NNG04GC89G. Based on $z$COSMOS observations carried out using the Very Large Telescope at the ESO Paranal Observatory under Programme ID: LP175.A-0839. Some of the data presented herein were obtained at the W.M. Keck Observatory, which is operated as a scientific partnership among the California Institute of Technology, the University of California and the National Aeronautics and Space Administration.

\begin{table}[h!tb!]
\begin{center}
\begin{tabular}{lll}
\#&Parameter&Description\\\hline
1 &  OBJNO   	&	Unique object number \\
2  & SURVEY\_ID	&	The unique survey ID, if available\\
3   &RA		&	Right Ascenion J2000 in decimal degrees\\
4  & DEC		&	Declination    J2000 in decimal degress\\
5   &NTOT\_HI	&	Total number of objects simultaniously fitted\\
6   &NTOT\_LOW&		Total number of objects simultaniously fitted	\\
7   &IMAGING	&	Imaging survey\\
8   &SPECZ 	&	Spectroscopic redshift\\
9  &PHOTOZ	&	Photometric redshift\\
10  &PHOTOZ\_CHI2	&	reduced chi2 for photometric redshift\\
11  &PHOTOZ\_ERR &		for EGS ($1\sigma$) and COSMOS (3$\sigma$)\\
12  &ZQUALITY	&	the quality flag for spectroscopic redshift\\
13  &Z\_ORIGIN&            Origin of Spectroscopic redshift      \\
14  &Z                          &      high quality specz else use photoz\\
15  &MAGB	&	B band apperant magnitude\\
16  &MAGB\_ERR	&	Error is B band magnitude\\
17  &MAGR	&	R band apperant magnitude\\
18  &MAGR\_ERR	&	Error in R band magnitude\\
19  &MAGI	&	I band apperant magnitude\\
20  &MAGI\_ERR&		Error in I band magnitude\\
21  &CFHT\_U    &          CFHTLS u mag\\
22  &CFHT\_U\_ERR  &        CFHTLS u mag error\\
23  &CFHT\_G           &   CFHTLS g mag\\
24  &CFHT\_G\_ERR  &        CFHTLS g mag error\\
25  &CFHT\_R          &    CFHTLS r mag\\
26  &CFHT\_R\_ERR  &        CFHTLS r mag error\\
27  &CFHT\_I           &   CFHTLS I mag\\
28  &CFHT\_I\_ERR  &        CFHTLS I mag error\\
29  &CFHT\_Z          &    CFHTLS z mag \\
30  &CFHT\_Z\_ERR  &        CFHTLS z mag error\\
31  &EBV              &     Extinction\\
32  &CLASS	&	Object classification, provided by DEEP2 survey and COMBO-17 survey\\
33  &MU\_HI   &            Surface Brightness    \\
34  &MU\_LOW    &          Surface Brightness  \\
35  &THETA\_IMAGE\_HI	&Theta image (SE\textsc{xtractor})\\
36  &THETA\_IMAGE\_LOW	&Theta image (SE\textsc{xtractor})\\
37  &THETA\_WORLD\_HI   	&Theta world (SE\textsc{xtractor})\\
38  &THETA\_WORLD\_LOW	&Theta world (SE\textsc{xtractor})\\
39  &BA\_HI	&	Axis ratio b/a (SE\textsc{xtractor})\\
40  &BA\_LOW	&	Axis ratio b/a (SE\textsc{xtractor})\\
\end{tabular}
\begin{center}
\end{center}
\end{center}
\end{table}

\begin{table}[h!tb!]
\begin{center}
\begin{tabular}{lll}

41  &KRON\_RADIUS\_HI&	Kron radius (SE\textsc{xtractor})\\
42  &KRON\_RADIUS\_LOW &	Kron radius (SE\textsc{xtractor})\\
43  &   FWHM\_HI & Full Width at Half Maximum ( SE\textsc{xtractor})\\        
44  &FWHM\_LOW  &   Full Width at Half Maximum ( SE\textsc{xtractor})   \\
45  &A\_IMAGE\_HI	&	A axis (SE\textsc{xtractor})\\
46  &A\_IMAGE\_LOW	&	A axis (SE\textsc{xtractor})\\
47  &B\_IMAGE\_HI	&	B axis (SE\textsc{xtractor})\\
48  &B\_IMAGE\_LOW &	B axis (SE\textsc{xtractor})\\
49  &BACKGROUND\_HI &	Sky background (SE\textsc{xtractor})\\
50  &BACKGROUND\_LOW	&Sky background (SE\textsc{xtractor})\\
51  &FLUX\_BEST\_HI &	Flux best (SE\textsc{xtractor})\\
52  &FLUX\_BEST\_LOW	&Flux best (SE\textsc{xtractor})\\
53  &FLUXERR\_BEST\_HI&	Error in flux best (SE\textsc{xtractor})\\
54  &FLUXERR\_BEST\_LOW&	Error in flux best (SE\textsc{xtractor})\\
55  &MAG\_BEST\_HI	&	Mag best (SE\textsc{xtractor})\\
56  &MAG\_BEST\_LOW	&Mag best (SE\textsc{xtractor})\\
57  &MAGERR\_BEST\_HI  &	Error in mag best (SE\textsc{xtractor})\\
58  &MAGERR\_BEST\_LOW	&Error in mag best (SE\textsc{xtractor})\\
59  &FLUX\_RADIUS\_HI	&Flux radius (SE\textsc{xtractor})\\
60  &FLUX\_RADIUS\_LOW	&Flux radius (SE\textsc{xtractor})\\
61  &ISOAREA\_IMAGE\_HI	&Iso area of object (SE\textsc{xtractor})\\
62  &ISOAREA\_IMAGE\_LOW	&Iso area of object (SE\textsc{xtractor})\\
63  &SEX\_FLAGS\_HI	&SE\textsc{xtractor} flag \\
64  &SEX\_FLAGS\_LOW 	&SE\textsc{xtractor} flag\\
65  &FLAG\_G\textsc{alfit}\_HI      &Flag G\textsc{alfit} good=0 bad=1\\   
66  &FLAG\_G\textsc{alfit}\_LOW&	Flag G\textsc{alfit} good=0 bad=1\\ 
67  &CHI2NU\_HI	&	G\textsc{alfit} reduced chi2\\
68  &CHI2NU\_LOW	&	G\textsc{alfit} reduced chi2\\
69  &CLASS\_STAR\_HI	&Class star (SE\textsc{xtractor}) \\
70  &CLASS\_STAR\_LOW&	Class star (SE\textsc{xtractor}) \\
71  &X\_G\textsc{alfit}\_HI      &   X center for G\textsc{alfit} residual image\\
72  &X\_G\textsc{alfit}\_LOW    &    X center for G\textsc{alfit} residual image\\
73  &Y\_G\textsc{alfit}\_HI        & Y center for G\textsc{alfit} residual image\\
74  &Y\_G\textsc{alfit}\_LOW      &  Y center for G\textsc{alfit} residual image\\
75  &MAG\_G\textsc{alfit}\_HI&	Mag (G\textsc{alfit})\\
76  &MAG\_G\textsc{alfit}\_LOW&	Mag (G\textsc{alfit})\\
77  &RE\_G\textsc{alfit}\_HI	 &Effective half-light radius (G\textsc{alfit})\\
78  &RE\_G\textsc{alfit}\_LOW	&Effective half-light radius (G\textsc{alfit})\\
79  &N\_G\textsc{alfit}\_HI	&	Sersic index [n $<$ 1.5 (Late type) n $>$ 2.5 (Early type)] (G\textsc{alfit})\\
80  &N\_G\textsc{alfit}\_LOW	&Sersic index [n $<$ 1.5 (Late type) n $>$ 2.5 (Early type)] (G\textsc{alfit})\\

\end{tabular}
\begin{center}
\end{center}
\end{center}
\end{table}

\begin{table*}[h!tb!]
\begin{center}
\begin{tabular}{lll}

81  &BA\_G\textsc{alfit}\_HI	 &Axis ratio (G\textsc{alfit})\\
82  &BA\_G\textsc{alfit}\_LOW&	Axis ratio (G\textsc{alfit})\\
83  &PA\_G\textsc{alfit}\_HI	&Position angle (G\textsc{alfit})\\
84  &PA\_G\textsc{alfit}\_LOW	&Position angle (G\textsc{alfit})  \\
85  &SKY\_G\textsc{alfit}\_HI	&Sky background measured by the G\textsc{alapagos} code\\
86  &SKY\_G\textsc{alfit}\_LOW	&Sky background measured by the G\textsc{alapagos} code\\
87  &MAGERR\_G\textsc{alfit}\_HI &	Error in mag (G\textsc{alfit})\\
88  &MAGERR\_G\textsc{alfit}\_LOW &	Error in mag (G\textsc{alfit})\\
89  &REERR\_G\textsc{alfit}\_HI&	Error in half-light radius (G\textsc{alfit})\\
90  &REERR\_G\textsc{alfit}\_LOW&	Error in half-light radius (G\textsc{alfit})\\
91 &NERR\_G\textsc{alfit}\_HI&	Error in sersic index (G\textsc{alfit})\\
92 &NERR\_G\textsc{alfit}\_LOW&	Error in sersic index (G\textsc{alfit})\\
93 &BAERR\_G\textsc{alfit}\_HI	&Error in axis ratio (G\textsc{alfit})\\
94 &BAERR\_G\textsc{alfit}\_LOW&	Error in axis ratio (G\textsc{alfit})\\
95 &PAERR\_G\textsc{alfit}\_HI&	Error in position angle (G\textsc{alfit}) \\
96 &PAERR\_G\textsc{alfit}\_LOW&	Error in position angle (G\textsc{alfit})\\
97& VIS\_MORPH	&	Visual morphology classification (currently not available)

\end{tabular}
\begin{center}
\end{center}
\end{center}
\end{table*}

\end{document}